\documentclass[11pt]{article}%
\usepackage{amsmath}
\usepackage{sw20jart}
\usepackage{epic}
\usepackage{amsfonts}
\usepackage{amssymb}
\usepackage{graphicx}%
\newtheorem{theorem}{Theorem}[section]

\newtheorem{lemma}[theorem]{Lemma}

\newtheorem{proposition}[theorem]{Proposition}

\ifx\pdfoutput\relax\let\pdfoutput=\undefined\fi
\newcount\msipdfoutput
\ifx\pdfoutput\undefined\else
\ifcase\pdfoutput\else
\ifx\paperwidth\undefined\else
\ifdim\paperheight=0pt\relax\else\pdfpageheight\paperheight\fi
\ifdim\paperwidth=0pt\relax\else\pdfpagewidth\paperwidth\fi
\fi\fi\fi
\begin{document}

\author{Mark Korenblit $^{1}$ and Vadim E. Levit $^{2}$\bigskip\\$^{1}$ Holon Institute of Technology, Israel\\korenblit@hit.ac.il\\$^{2}$ Ariel University Center of Samaria, Israel\\levitv@ariel.ac.il }
\title{A One-Vertex Decomposition Algorithm for Generating Algebraic Expressions of
Square Rhomboids }
\date{}
\maketitle

\begin{abstract}
The paper investigates relationship between algebraic expressions and graphs.
We consider a digraph called a square rhomboid that is an example of
non-series-parallel graphs. Our intention is to simplify the expressions of
square rhomboids and eventually find their shortest representations. With that
end in view, we describe the new algorithm for generating square rhomboid
expressions, which improves on our previous algorithms.

\end{abstract}

\section{Introduction\label{intro}}

A \textit{graph }$G=(V,E)$ consists of a \textit{vertex set\ }$V$ and an
\textit{edge set\ }$E$, where each edge corresponds to a pair $(v,w)$ of
vertices. A graph in which each edge has a \textit{weight} associated with it
is called a \textit{network}. A graph $G^{\prime}=(V^{\prime},E^{\prime})$ is
a \textit{subgraph} of $G=(V,E)$ if $V^{\prime}\subseteq V$ and $E^{\prime
}\subseteq E$. A graph $G$ is a \textit{homeomorph} of $G^{\prime}$ if $G$ can
be obtained by subdividing edges of $G^{\prime}$ with new vertices. We say
that a graph $G^{2}=(V,E^{\prime})$ is a \textit{square of a graph} $G=(V,E)$
if $E^{\prime}=\left\{  (u,w):(u,w)\in E\vee\left(  (u,v)\in E\wedge(v,w)\in
E\right)  \text{ for some }v\in V\right\}  $. A two-terminal directed acyclic
graph (\textit{st-dag}) has only one source and only one sink.

We consider a \textit{labeled graph} which has labels attached to its edges.
Each path between the source and the sink (a \textit{sequential path}) in an
st-dag can be presented by a product of all edge labels of the path. We define
the sum of edge label products corresponding to all possible sequential paths
of an st-dag $G$ as the \textit{canonical expression }of $G$. An algebraic
expression is called an \textit{st-dag expression} (a \textit{factoring of an
st-dag} in \cite{BKS}) if it is algebraically equivalent to the canonical
expression of an st-dag. An st-dag expression consists of literals (edge
labels), and the operators $+$ (disjoint union) and $\cdot$ (concatenation,
also denoted by juxtaposition). An expression of an st-dag $G$ will be
hereafter denoted by $Ex(G)$.

We define the total number of literals in an algebraic expression as the
\textit{complexity of the algebraic expression}. An equivalent expression with
the minimum complexity is called an \textit{optimal representation of the
algebraic expression}.

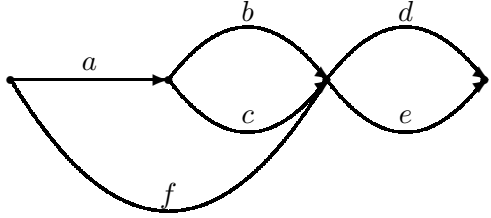
\begin{figure}[ptb]
\setlength{\unitlength}{0.7cm}
\par
\begin{picture}(5,3.5)(-5,0)\thicklines
\multiput(1,3)(3,0){4}{\circle*{0.15}}
\put(1,3){\vector(1,0){3}} \put(2.5,3.3){\makebox(0,0){$a$}}
\qbezier(4,3)(5.5,5)(7,3) \put(7.085,3){\vector(3,-2){0}}
\put(5.5,4.3){\makebox(0,0){$b$}}
\qbezier(4,3)(5.5,1)(7,3) \put(7.085,3){\vector(3,2){0}}
\put(5.5,2.3){\makebox(0,0){$c$}}
\qbezier(7,3)(8.5,5)(10,3) \put(10.085,3){\vector(3,-2){0}}
\put(8.5,4.3){\makebox(0,0){$d$}}
\qbezier(7,3)(8.5,1)(10,3) \put(10.085,3){\vector(3,2){0}}
\put(8.5,2.3){\makebox(0,0){$e$}}
\qbezier(1,3)(4,-2)(7,3) \put(7.085,3){\vector(4,3){0}}
\put(4,0.8){\makebox(0,0){$f$}}
\end{picture}\caption{A series-parallel graph.}%
\label{fig1}%
\end{figure}

A \textit{series-parallel} \textit{graph} is defined recursively so that a
single edge is a series-parallel graph and a graph obtained by a parallel or a
series composition of series-parallel graphs is series-parallel. As shown in
\cite{BKS} and \cite{KoL}, a series-parallel graph expression has a
representation in which each literal appears only once. This representation is
an optimal representation of the series-parallel graph expression. For
example, the canonical expression of the series-parallel graph presented in
Figure \ref{fig1} is $abd+abe+acd+ace+fe+fd$. Since it is a series-parallel
graph, the expression can be reduced to $(a(b+c)+f)(d+e)$, where each literal
appears once.

A \textit{Fibonacci graph} \cite{GoP} has vertices $\{1,2,3,\ldots,n\}$ and
edges $\{\left(  v,v+1\right)  \mid v=\nolinebreak1,2,\ldots,$\newline%
$n-1\}\cup\left\{  \left(  v,v+2\right)  \mid v=1,2,\ldots,n-2\right\}  $. As
shown in \cite{Duf}, an st-dag is series-parallel if and only if it does not
contain a subgraph which is a homeomorph of the \textit{forbidden subgraph}
positioned between vertices $1$ and $4$ of the Fibonacci graph illustrated in
Figure \ref{fig2}. Thus, a Fibonacci graph gives a generic example of
non-series-parallel graphs.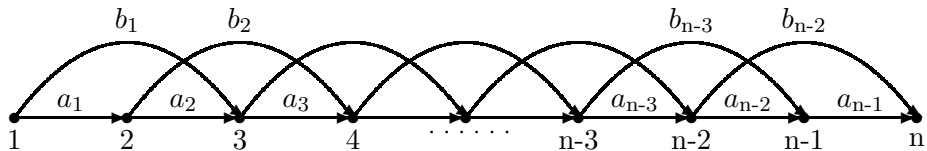
\begin{figure}[ptbh]
\setlength{\unitlength}{1.0cm}
\par
\begin{picture}(5,2)(-0.9,-0.5)\thicklines
\multiput(0,0)(1.5,0){9}{\circle*{0.15}}
\put(0,-0.3){\makebox(0,0){1}}
\put(1.5,-0.3){\makebox(0,0){2}}
\put(3,-0.3){\makebox(0,0){3}}
\put(4.5,-0.3){\makebox(0,0){4}}
\put(7.5,-0.3){\makebox(0,0){n-3}}
\put(9,-0.3){\makebox(0,0){n-2}}
\put(10.5,-0.3){\makebox(0,0){n-1}}
\put(12,-0.3){\makebox(0,0){n}}
\multiput(0,0)(1.5,0){8}{\vector(1,0){1.5}}
\put(0.75,0.2){\makebox(0,0){$a_{1}$}}
\put(2.25,0.2){\makebox(0,0){$a_{2}$}}
\put(3.75,0.2){\makebox(0,0){$a_{3}$}}
\put(8.25,0.2){\makebox(0,0){$a_{\text{n-3}}$}}
\put(9.75,0.2){\makebox(0,0){$a_{\text{n-2}}$}}
\put(11.25,0.2){\makebox(0,0){$a_{\text{n-1}}$}}
\qbezier(0,0)(1.5,2)(3,0)
\qbezier(1.5,0)(3,2)(4.5,0)
\qbezier(3,0)(4.5,2)(6,0)
\qbezier(4.5,0)(6,2)(7.5,0)
\qbezier(6,0)(7.5,2)(9,0)
\qbezier(7.5,0)(9,2)(10.5,0)
\qbezier(9,0)(10.5,2)(12,0)
\multiput(3.085,0)(1.5,0){7}{\vector(3,-2){0}}
\put(1.5,1.3){\makebox(0,0){$b_{1}$}}
\put(3,1.3){\makebox(0,0){$b_{2}$}}
\put(9,1.3){\makebox(0,0){$b_{\text{n-3}}$}}
\put(10.5,1.3){\makebox(0,0){$b_{\text{n-2}}$}}
\multiput(5.55,-0.2)(0.2,0){6}{\circle*{0.02}}
\end{picture}
\caption{A Fibonacci graph.}%
\label{fig2}%
\end{figure}

Mutual relations between graphs and expressions are discussed in a number of
works. Specifically, \cite{Mun1}, \cite{Mun2}, and \cite{SaW} consider the
correspondence between series-parallel graphs and read-once functions. A
Boolean function is defined as \textit{read-once} if it may be computed by
some formula in which no variable occurs more than once (\textit{read-once
formula}). On the other hand, a series-parallel graph expression can be
reduced to the representation in which each literal appears only once. Hence,
such a representation of a series-parallel graph expression can be considered
as a read-once formula (boolean operations are replaced by arithmetic ones).

Problems related to computations on graphs have applications in different areas.

Specifically, many network problems, which are either intractable or have
complicated solutions in the general case are solvable for series-parallel
graphs. For example, some efficient algorithms for flow problems on
series-parallel networks are presented in \cite{BeB}, \cite{BBT}, \cite{JaC},
\cite{Tam}. Papers \cite{AbK}, \cite{FLMB}, \cite{Mon}, \cite{MoS} consider
sequencing and scheduling in relation to precedence series-parallel
constraints. Linear algorithms for reliability problems on series-parallel
networks are presented in \cite{SatW}, \cite{WaC}.

An expression of a homeomorph of the forbidden subgraph belonging to any
non-series-parallel st-dag has no representation in which each literal appears
once. For example, consider the subgraph positioned between vertices $1$ and
$4$ of the Fibonacci graph shown in Figure \ref{fig2}. Possible optimal
representations of its expression are $a_{1}\left(  a_{2}a_{3}+b_{2}\right)
+b_{1}a_{3}$ or $\left(  a_{1}a_{2}+b_{1}\right)  a_{3}+a_{1}b_{2}$. For this
reason, an expression of a non-series-parallel st-dag can not be represented
as a read-once formula. However, for arbitrary functions, which are not
read-once, generating the optimum factored form is NP-complete \cite{Wan}.

The problem of factoring boolean functions into shorter, more compact formulae
is one of the basic operations in algorithmic logic synthesis. In logic
synthesis, one standard measure of the complexity of a logic circuit is the
number of literals. Computation time also depends on the number of literals.
Some algorithms developed in order to obtain good factored forms are described
in \cite{GoM}, \cite{GMR}.

A symbolic approach to scheduling of a robotic line is considered in
\cite{LeK}. The method uses the max-algebra tools and allows the shortest-path
problem to be interpreted as the computation of the st-dag expression. The
complexity of this problem is determined by the complexity of the st-dag
expression. For a robotic line represented by a Fibonacci graph, the proposed
algorithm generates the processing sequence in polynomial time.

A method for automated composition of algebraic expressions in complex
business process modeling based on acyclic directed graph reductions is
introduced in \cite{OFMP}. The method transforms business step dependencies
described by users into digraphs and finally generates algebraic expressions.
If a graph is not series-parallel, the algorithm checks potential structural
conflicts whose presence complicates certain aspects, such as execution
control and system scalability. In this case, the expression generation may
require exponential time.

A lot of problems have polynomial algorithms on \textit{planar graphs}, i.e.,
the graphs that can be drawn in the plane without any edges crossing. For
example, some polynomial-time algorithms for the exact computation of network
reliability applied to planar graphs are surveyed in \cite{PoS}. In \cite{PrB}
a polynomial algorithm for the connectivity augmentation problem is presented.
This problem has applications in upgrading telecommunication networks to be
invulnerable to link or node failures. There are also many practical
applications with a graph structure in which crossing edges are a nuisance,
including design problems for circuits, subways, utility lines \cite{Fle}.

In our work we consider expressions with a minimum (or, at least, a
polynomial) complexity as a key to generating efficient algorithms on
distributed systems.

In \cite{KoL} we presented an algorithm,\ which generates the expression of
$O\left(  n^{2}\right)  $ complexity for an $n$-vertex Fibonacci graph.

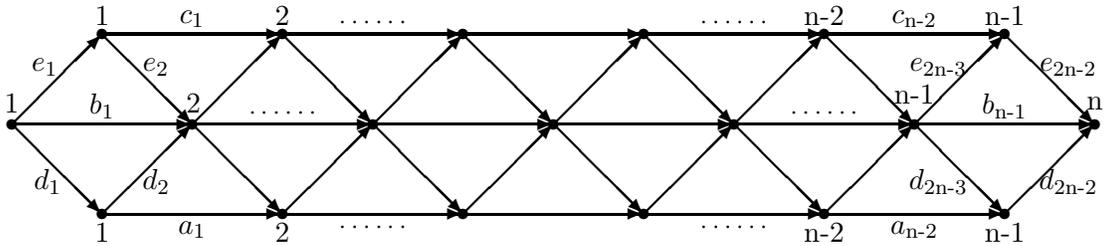
\begin{figure}[ptbh]
\setlength{\unitlength}{0.8cm}
\par
\begin{picture}(10,4)(-0.2,-1.7)\thicklines
\multiput(0,0)(3,0){7}{\circle*{0.1875}}
\multiput(1.5,1.5)(3,0){6}{\circle*{0.1875}}
\multiput(1.5,-1.5)(3,0){6}{\circle*{0.1875}}
\put(0,0.35){\makebox(0,0){1}} \put(3.02,0.35){\makebox(0,0){2}}
\put(15,0.5){\makebox(0,0){n-1}} \put(18,0.35){\makebox(0,0){n}}
\put(1.5,1.8){\makebox(0,0){1}} \put(4.5,1.8){\makebox(0,0){2}}
\put(13.5,1.8){\makebox(0,0){n-2}}
\put(16.5,1.8){\makebox(0,0){n-1}}
\put(1.5,-1.8){\makebox(0,0){1}} \put(4.5,-1.8){\makebox(0,0){2}}
\put(13.5,-1.8){\makebox(0,0){n-2}}
\put(16.5,-1.8){\makebox(0,0){n-1}}
\multiput(0,0)(3,0){6}{\vector(1,1){1.5}}
\multiput(0,0)(3,0){6}{\vector(1,-1){1.5}}
\multiput(1.5,1.5)(3,0){6}{\vector(1,-1){1.5}}
\multiput(1.5,-1.5)(3,0){6}{\vector(1,1){1.5}}
\multiput(0,0)(3,0){6}{\vector(1,0){3}}
\multiput(1.5,1.5)(3,0){5}{\vector(1,0){3}}
\multiput(1.5,-1.5)(3,0){5}{\vector(1,0){3}}
\put(1.5,0.3){\makebox(0,0){$b_{1}$}}
\put(16.5,0.3){\makebox(0,0){$b_{\text{n-1}}$}}
\put(3,1.75){\makebox(0,0){$c_{1}$}}
\put(15,1.75){\makebox(0,0){$c_{\text{n-2}}$}}
\put(3,-1.75){\makebox(0,0){$a_{1}$}}
\put(15,-1.75){\makebox(0,0){$a_{\text{n-2}}$}}
\put(0.55,0.95){\makebox(0,0){$e_{1}$}}
\put(15.4,0.95){\makebox(0,0){$e_{\text{2n-3}}$}}
\put(2.4,0.95){\makebox(0,0){$e_{2}$}}
\put(17.56,0.95){\makebox(0,0){$e_{\text{2n-2}}$}}
\put(0.6,-0.95){\makebox(0,0){$d_{1}$}}
\put(15.4,-0.95){\makebox(0,0){$d_{\text{2n-3}}$}}
\put(2.4,-0.95){\makebox(0,0){$d_{2}$}}
\put(17.56,-0.95){\makebox(0,0){$d_{\text{2n-2}}$}}
\multiput(4,0.2)(0.2,0){6}{\circle*{0.02}}
\multiput(13,0.2)(0.2,0){6}{\circle*{0.02}}
\multiput(5.5,1.7)(0.2,0){6}{\circle*{0.02}}
\multiput(11.5,1.7)(0.2,0){6}{\circle*{0.02}}
\multiput(5.5,-1.7)(0.2,0){6}{\circle*{0.02}}
\multiput(11.5,-1.7)(0.2,0){6}{\circle*{0.02}}
\end{picture}\caption{A square rhomboid of size $n$.}%
\label{rhom_fig12}%
\end{figure}

In this paper we investigate a non-series-parallel st-dag called a
\textit{square rhomboid} (Figure \ref{rhom_fig12}). This graph looks like a
planar approximation of the square of a \textit{rhomboid}, which is a series
composition of \textit{rhomb} graphs. A square rhomboid consists of the same
vertices as the corresponding rhomboid. However, edges labeled by letters $a$,
$b$, and $c$ (see Figure \ref{rhom_fig12}) are absent in a rhomboid.
Geometrically, a square rhomboid can be considered to be a \textquotedblright
gluing\textquotedblright\ of two Fibonacci graphs, i.e., it is the next harder
one in a sequence of increasingly non-series-parallel graphs.

The set of vertices of an $N$-vertex square rhomboid consists of $\frac
{N+2}{3}$ \textit{middle} (\textit{basic}), $\frac{N-1}{3}$ \textit{upper},
and\textit{\ }$\frac{N-1}{3}$ \textit{lower} vertices. Upper and lower
vertices numbered $x$ will be denoted in formulae by $\overline{x}$ and
$\underline{x} $, respectively. The square rhomboid ($SR$ for brevity)
including $n$ basic vertices will be denoted by $SR(n)$ and will be called an
$SR$ of size $n$.

Our intention in this paper is to simplify the expressions of square rhomboids
and eventually find their optimal representations. With that end in view, we
present an algorithm based on a \textit{decomposition method}.

\section{A One-Vertex Decomposition Method (1-VDM)}

The method is based on revealing subgraphs in the initial graph. The resulting
expression is produced by a special composition of subexpressions describing
these subgraphs.

\begin{figure}[t]
\setlength{\unitlength}{0.7cm}
\par
\begin{picture}(10,4)(-1.1,-2.5)\thicklines
\multiput(0,0)(3,0){7}{\circle*{0.1875}}
\multiput(1.5,1.5)(3,0){6}{\circle*{0.1875}}
\multiput(1.5,-1.5)(3,0){6}{\circle*{0.1875}}
\put(0,0.35){\makebox(0,0){1}} \put(3.02,0.35){\makebox(0,0){2}}
\put(6,0.35){\makebox(0,0){3}} \put(9,0.35){\makebox(0,0){4}}
\put(12,0.35){\makebox(0,0){5}} \put(15,0.35){\makebox(0,0){6}}
\put(18,0.35){\makebox(0,0){7}}
\put(1.5,1.8){\makebox(0,0){1}} \put(4.5,1.8){\makebox(0,0){2}}
\put(7.5,1.8){\makebox(0,0){3}} \put(10.5,1.8){\makebox(0,0){4}}
\put(13.5,1.8){\makebox(0,0){5}} \put(16.5,1.8){\makebox(0,0){6}}
\put(1.5,-1.8){\makebox(0,0){1}} \put(4.5,-1.8){\makebox(0,0){2}}
\put(7.5,-1.8){\makebox(0,0){3}} \put(10.5,-1.8){\makebox(0,0){4}}
\put(13.5,-1.8){\makebox(0,0){5}} \put(16.5,-1.8){\makebox(0,0){6}}
\multiput(0,0)(3,0){6}{\vector(1,1){1.5}}
\multiput(0,0)(3,0){6}{\vector(1,-1){1.5}}
\multiput(1.5,1.5)(3,0){6}{\vector(1,-1){1.5}}
\multiput(1.5,-1.5)(3,0){6}{\vector(1,1){1.5}}
\multiput(0,0)(3,0){6}{\vector(1,0){3}}
\multiput(1.5,1.5)(3,0){5}{\vector(1,0){3}}
\multiput(1.5,-1.5)(3,0){5}{\vector(1,0){3}}
\put(1.5,0.3){\makebox(0,0){$b_{1}$}}
\put(4.5,0.3){\makebox(0,0){$b_{2}$}}
\put(7.5,0.3){\makebox(0,0){$b_{3}$}}
\put(10.5,0.3){\makebox(0,0){$b_{4}$}}
\put(13.5,0.3){\makebox(0,0){$b_{5}$}}
\put(16.5,0.3){\makebox(0,0){$b_{6}$}}
\put(3,1.75){\makebox(0,0){$c_{1}$}}
\put(6,1.75){\makebox(0,0){$c_{2}$}}
\put(8.7,1.75){\makebox(0,0){$c_{3}$}}
\put(12,1.75){\makebox(0,0){$c_{4}$}}
\put(15,1.75){\makebox(0,0){$c_{5}$}}
\put(3,-1.75){\makebox(0,0){$a_{1}$}}
\put(6,-1.75){\makebox(0,0){$a_{2}$}}
\put(8.7,-1.75){\makebox(0,0){$a_{3}$}}
\put(12,-1.75){\makebox(0,0){$a_{4}$}}
\put(15,-1.75){\makebox(0,0){$a_{5}$}}
\put(0.55,0.95){\makebox(0,0){$e_{1}$}}
\put(3.55,0.95){\makebox(0,0){$e_{3}$}}
\put(6.55,0.95){\makebox(0,0){$e_{5}$}}
\put(9.55,0.95){\makebox(0,0){$e_{7}$}}
\put(12.55,0.95){\makebox(0,0){$e_{9}$}}
\put(15.5,0.95){\makebox(0,0){$e_{11}$}}
\put(2.4,0.95){\makebox(0,0){$e_{2}$}}
\put(5.4,0.95){\makebox(0,0){$e_{4}$}}
\put(8.4,0.95){\makebox(0,0){$e_{6}$}}
\put(11.4,0.95){\makebox(0,0){$e_{8}$}}
\put(14.45,0.95){\makebox(0,0){$e_{10}$}}
\put(17.45,0.95){\makebox(0,0){$e_{12}$}}
\put(0.6,-0.95){\makebox(0,0){$d_{1}$}}
\put(3.6,-0.95){\makebox(0,0){$d_{3}$}}
\put(6.6,-0.95){\makebox(0,0){$d_{5}$}}
\put(9.6,-0.95){\makebox(0,0){$d_{7}$}}
\put(12.6,-0.95){\makebox(0,0){$d_{9}$}}
\put(15.6,-0.95){\makebox(0,0){$d_{11}$}}
\put(2.4,-0.95){\makebox(0,0){$d_{2}$}}
\put(5.4,-0.95){\makebox(0,0){$d_{4}$}}
\put(8.4,-0.95){\makebox(0,0){$d_{6}$}}
\put(11.4,-0.95){\makebox(0,0){$d_{8}$}}
\put(14.45,-0.95){\makebox(0,0){$d_{10}$}}
\put(17.45,-0.95){\makebox(0,0){$d_{12}$}}
\dashline[75]{0.15}(9,2.5)(9,0.55)
\dashline[75]{0.15}(9,0.1)(9,-2.5)
\end{picture}\caption{Decomposition of a square rhomboid of size $7$ at vertex
$4$.}%
\label{rhom_fig3}%
\end{figure}
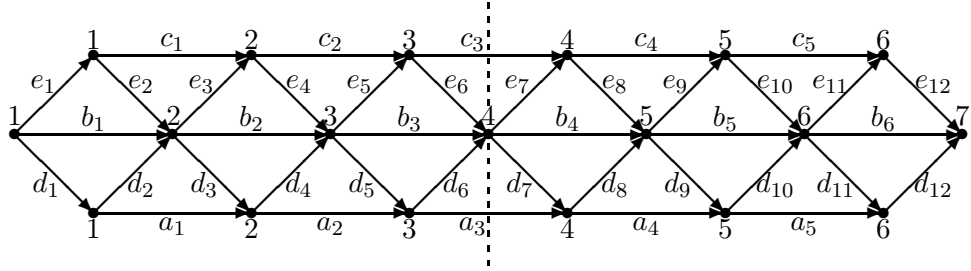

For a non-trivial $SR$ subgraph with a source $p$ and a sink $q$ we choose any
\textit{decomposition vertex} $i$ ($p<i<q$) located in the basic group of a
subgraph. We conditionally split each $SR$ through its decomposition vertex
(see the example in Figure \ref{rhom_fig3}).\pagebreak

Two kinds of subgraphs are revealed in the graph in the course of
decomposition. The first of them is an $SR$ with a fewer number of vertices
than the initial $SR$. The second one is an $SR$ supplemented by two
additional edges at one of four sides. Possible varieties of this st-dag (we
call it a\textit{\ single-leaf square rhomboid} and denote by $\widehat{SR}$)
which are subgraphs revealed from an $SR$ in Figure \ref{rhom_fig3}, are
illustrated in Figure \ref{rhom_fig4}(a, b, c, d). Let $\widehat{SR}(n)$ (an
$\widehat{SR}$ of size $n$) denote an $\widehat{SR}$ including $n$ basic vertices.

We denote by $E(p,q)$ a subexpression related to an $SR$ subgraph with a
source $p$ and a sink $q$. We denote by $E(p,\overline{q})$, $E(\overline
{p},q)$, $E(p,\underline{q})$, $E(\underline{p},q)$ subexpressions related to
$\widehat{SR}$ subgraphs of Figure \ref{rhom_fig4}(a, b, c, d, respectively)
with a source $p$ and a sink $\overline{q}$, a source $\overline{p}$ and a
sink $q$, a source $p$ and a sink $\underline{q}$, and a source $\underline
{p}$ and a sink $q$.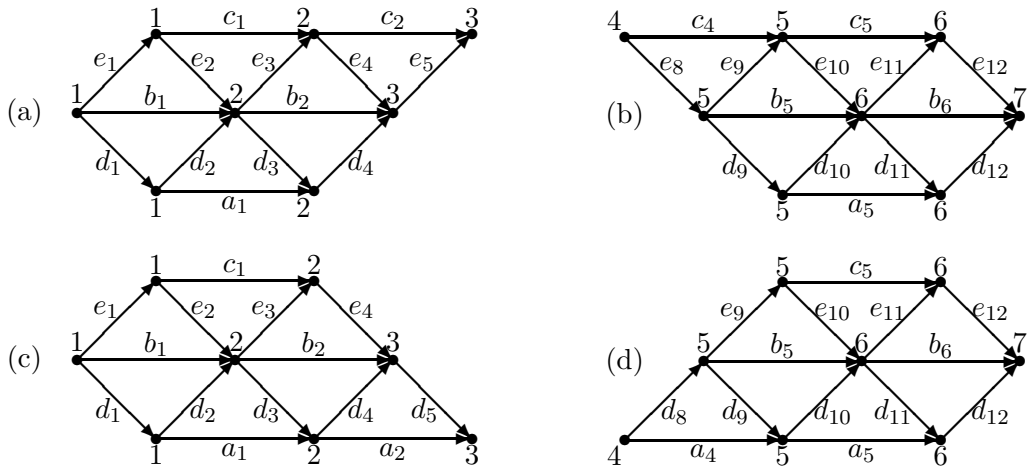
\begin{figure}[ptbh]
\setlength{\unitlength}{0.7cm}
\par
\begin{picture}(10,3)(-1.3,-1)\thicklines
\put(-1,0){\makebox(0,0){(a)}}
\multiput(0,0)(3,0){3}{\circle*{0.1875}}
\multiput(1.5,1.5)(3,0){3}{\circle*{0.1875}}
\multiput(1.5,-1.5)(3,0){2}{\circle*{0.1875}}
\put(0,0.35){\makebox(0,0){1}}
\put(3.02,0.35){\makebox(0,0){2}}
\put(6,0.35){\makebox(0,0){3}}
\put(1.5,1.8){\makebox(0,0){1}}
\put(4.3,1.8){\makebox(0,0){2}}
\put(7.5,1.8){\makebox(0,0){3}}
\put(1.5,-1.8){\makebox(0,0){1}}
\put(4.3,-1.8){\makebox(0,0){2}}
\multiput(0,0)(3,0){3}{\vector(1,1){1.5}}
\multiput(0,0)(3,0){2}{\vector(1,-1){1.5}}
\multiput(1.5,1.5)(3,0){2}{\vector(1,-1){1.5}}
\multiput(1.5,-1.5)(3,0){2}{\vector(1,1){1.5}}
\multiput(0,0)(3,0){2}{\vector(1,0){3}}
\multiput(1.5,1.5)(3,0){2}{\vector(1,0){3}}
\multiput(1.5,-1.5)(3,0){1}{\vector(1,0){3}}
\put(1.5,0.3){\makebox(0,0){$b_{1}$}}
\put(4.2,0.3){\makebox(0,0){$b_{2}$}}
\put(3,1.75){\makebox(0,0){$c_{1}$}}
\put(6,1.75){\makebox(0,0){$c_{2}$}}
\put(3,-1.75){\makebox(0,0){$a_{1}$}}
\put(0.55,0.95){\makebox(0,0){$e_{1}$}}
\put(3.55,0.95){\makebox(0,0){$e_{3}$}}
\put(6.55,0.95){\makebox(0,0){$e_{5}$}}
\put(2.4,0.95){\makebox(0,0){$e_{2}$}}
\put(5.4,0.95){\makebox(0,0){$e_{4}$}}
\put(0.6,-0.95){\makebox(0,0){$d_{1}$}}
\put(3.6,-0.95){\makebox(0,0){$d_{3}$}}
\put(2.4,-0.95){\makebox(0,0){$d_{2}$}}
\put(5.4,-0.95){\makebox(0,0){$d_{4}$}}
\end{picture}
\par
\begin{picture}(10,4.5)(-1.3,-0.85)\thicklines
\put(-1,0){\makebox(0,0){(c)}}
\multiput(0,0)(3,0){3}{\circle*{0.1875}}
\multiput(1.5,1.5)(3,0){2}{\circle*{0.1875}}
\multiput(1.5,-1.5)(3,0){3}{\circle*{0.1875}}
\put(0,0.35){\makebox(0,0){1}}
\put(3.02,0.35){\makebox(0,0){2}}
\put(6,0.35){\makebox(0,0){3}}
\put(1.5,1.8){\makebox(0,0){1}}
\put(4.5,1.8){\makebox(0,0){2}}
\put(1.5,-1.8){\makebox(0,0){1}}
\put(4.5,-1.8){\makebox(0,0){2}}
\put(7.5,-1.8){\makebox(0,0){3}}
\multiput(0,0)(3,0){2}{\vector(1,1){1.5}}
\multiput(0,0)(3,0){3}{\vector(1,-1){1.5}}
\multiput(1.5,1.5)(3,0){2}{\vector(1,-1){1.5}}
\multiput(1.5,-1.5)(3,0){2}{\vector(1,1){1.5}}
\multiput(0,0)(3,0){2}{\vector(1,0){3}}
\multiput(1.5,1.5)(3,0){1}{\vector(1,0){3}}
\multiput(1.5,-1.5)(3,0){2}{\vector(1,0){3}}
\put(1.5,0.3){\makebox(0,0){$b_{1}$}}
\put(4.5,0.3){\makebox(0,0){$b_{2}$}}
\put(3,1.75){\makebox(0,0){$c_{1}$}}
\put(3,-1.75){\makebox(0,0){$a_{1}$}}
\put(6,-1.75){\makebox(0,0){$a_{2}$}}
\put(0.55,0.95){\makebox(0,0){$e_{1}$}}
\put(3.55,0.95){\makebox(0,0){$e_{3}$}}
\put(2.4,0.95){\makebox(0,0){$e_{2}$}}
\put(5.4,0.95){\makebox(0,0){$e_{4}$}}
\put(0.6,-0.95){\makebox(0,0){$d_{1}$}}
\put(3.6,-0.95){\makebox(0,0){$d_{3}$}}
\put(6.6,-0.95){\makebox(0,0){$d_{5}$}}
\put(2.4,-0.95){\makebox(0,0){$d_{2}$}}
\put(5.4,-0.95){\makebox(0,0){$d_{4}$}}
\end{picture}
\par
\begin{picture}(10,0)(-1.2,-1.5)\thicklines
\put(10.5,0){\makebox(0,0){(d)}}
\multiput(12,0)(3,0){3}{\circle*{0.1875}}
\multiput(13.5,1.5)(3,0){2}{\circle*{0.1875}}
\multiput(10.5,-1.5)(3,0){3}{\circle*{0.1875}}
\put(12,0.35){\makebox(0,0){5}}
\put(15,0.35){\makebox(0,0){6}}
\put(18,0.35){\makebox(0,0){7}}
\put(13.5,1.8){\makebox(0,0){5}}
\put(16.5,1.8){\makebox(0,0){6}}
\put(10.3,-1.8){\makebox(0,0){4}}
\put(13.5,-1.8){\makebox(0,0){5}}
\put(16.5,-1.8){\makebox(0,0){6}}
\multiput(12,0)(3,0){2}{\vector(1,1){1.5}}
\multiput(12,0)(3,0){2}{\vector(1,-1){1.5}}
\multiput(13.5,1.5)(3,0){2}{\vector(1,-1){1.5}}
\multiput(10.5,-1.5)(3,0){3}{\vector(1,1){1.5}}
\multiput(12,0)(3,0){2}{\vector(1,0){3}}
\multiput(13.5,1.5)(3,0){1}{\vector(1,0){3}}
\multiput(10.5,-1.5)(3,0){2}{\vector(1,0){3}}
\put(13.5,0.3){\makebox(0,0){$b_{5}$}}
\put(16.5,0.3){\makebox(0,0){$b_{6}$}}
\put(15,1.75){\makebox(0,0){$c_{5}$}}
\put(12,-1.75){\makebox(0,0){$a_{4}$}}
\put(15,-1.75){\makebox(0,0){$a_{5}$}}
\put(12.55,0.95){\makebox(0,0){$e_{9}$}}
\put(15.5,0.95){\makebox(0,0){$e_{11}$}}
\put(14.45,0.95){\makebox(0,0){$e_{10}$}}
\put(17.45,0.95){\makebox(0,0){$e_{12}$}}
\put(12.6,-0.95){\makebox(0,0){$d_{9}$}}
\put(15.6,-0.95){\makebox(0,0){$d_{11}$}}
\put(11.45,-0.95){\makebox(0,0){$d_{8}$}}
\put(14.45,-0.95){\makebox(0,0){$d_{10}$}}
\put(17.45,-0.95){\makebox(0,0){$d_{12}$}}
\end{picture}
\par
\begin{picture}(10,0)(-1.2,-6.85)\thicklines
\put(10.5,0){\makebox(0,0){(b)}}
\multiput(12,0)(3,0){3}{\circle*{0.1875}}
\multiput(10.5,1.5)(3,0){3}{\circle*{0.1875}}
\multiput(13.5,-1.5)(3,0){2}{\circle*{0.1875}}
\put(12,0.35){\makebox(0,0){5}} \put(15,0.35){\makebox(0,0){6}}
\put(18,0.35){\makebox(0,0){7}}
\put(10.3,1.8){\makebox(0,0){4}} \put(13.5,1.8){\makebox(0,0){5}}
\put(16.5,1.8){\makebox(0,0){6}}
\put(13.5,-1.8){\makebox(0,0){5}} \put(16.5,-1.8){\makebox(0,0){6}}
\multiput(12,0)(3,0){2}{\vector(1,1){1.5}}
\multiput(12,0)(3,0){2}{\vector(1,-1){1.5}}
\multiput(10.5,1.5)(3,0){3}{\vector(1,-1){1.5}}
\multiput(13.5,-1.5)(3,0){2}{\vector(1,1){1.5}}
\multiput(12,0)(3,0){2}{\vector(1,0){3}}
\multiput(10.5,1.5)(3,0){2}{\vector(1,0){3}}
\multiput(13.5,-1.5)(3,0){1}{\vector(1,0){3}}
\put(13.5,0.3){\makebox(0,0){$b_{5}$}}
\put(16.5,0.3){\makebox(0,0){$b_{6}$}}
\put(12,1.75){\makebox(0,0){$c_{4}$}}
\put(15,1.75){\makebox(0,0){$c_{5}$}}
\put(15,-1.75){\makebox(0,0){$a_{5}$}}
\put(12.55,0.95){\makebox(0,0){$e_{9}$}}
\put(15.5,0.95){\makebox(0,0){$e_{11}$}}
\put(11.4,0.95){\makebox(0,0){$e_{8}$}}
\put(14.45,0.95){\makebox(0,0){$e_{10}$}}
\put(17.45,0.95){\makebox(0,0){$e_{12}$}}
\put(12.6,-0.95){\makebox(0,0){$d_{9}$}}
\put(15.6,-0.95){\makebox(0,0){$d_{11}$}}
\put(14.45,-0.95){\makebox(0,0){$d_{10}$}}
\put(17.45,-0.95){\makebox(0,0){$d_{12}$}}
\end{picture}\caption{Varieties of a single-leaf square rhomboid of size $3$.}%
\label{rhom_fig4}%
\end{figure}

Any path from vertex $1$ to vertex $7$ in Fig. \ref{rhom_fig3} passes through
decomposition vertex $4$ or through edge $c_{3}$ or through edge $a_{3}$.
Therefore, $E(p,q)$ is generated by the following recursive procedure
(\textit{decomposition procedure}):

\begin{enumerate}
\item \label{mr1}$\mathbf{case}$ $q=p:E(p,q)\leftarrow1$

\item \label{mr2}$\mathbf{case}$ $q=p+1:E(p,q)\leftarrow b_{p}+e_{2p-1}%
e_{2p}+d_{2p-1}d_{2p}$

\item \label{mr3}$\mathbf{case}$ $q>p+1:\mathbf{choice}(p,q,i)$

\item \label{mr4}$\qquad\qquad\qquad\quad~E(p,q)\leftarrow
E(p,i)E(i,q)+E(p,\overline{i-1})c_{i-1}E(\overline{i},q)+$

$\qquad\qquad\qquad\qquad\qquad\qquad E(p,\underline{i-1})a_{i-1}%
E(\underline{i},q)\qquad$
\end{enumerate}

The expression related to a one-vertex $SR$ is defined formally as $1$ (line
\ref{mr1}). Line \ref{mr2} describes the expression of $SR(2)$ which is a
trivial subgraph. The procedure $\mathbf{choice}(p,q,i)$ in line \ref{mr3}
chooses an arbitrary number $i$ on the interval $(p,q)$ so that $p<i<q$. This
number is a number of a decomposition vertex in a basic group of the $SR$
subgraph positioned between vertices $p$ and $q$. A current subgraph is
decomposed into six new subgraphs in line \ref{mr4}. Subgraphs described by
subexpressions $E(p,i)$ and $E(i,q)$ include all paths from vertex $p$ to
vertex $q$ passing through vertex $i$. Subgraphs described by subexpressions
$E(p,\overline{i-1})$ and $E(\overline{i},q)$ include all paths from vertex
$p$ to vertex $q$ passing via edge $c_{i-1}$. Subgraphs described by
subexpressions $E(p,\underline{i-1})$ and $E(\underline{i},q)$ include all
paths from vertex $p$ to vertex $q$ passing via edge $a_{i-1}$.

$E(1,n)$ is the expression of the initial $SR$ of size $n$. Hence, the
decomposition procedure is initially invoked by substituting $1$ and $n$
instead of $p$ and $q$, respectively.

We now consider the algorithm that generates the algebraic expression for a
square rhomboid using 1-VDM as its base.

\section{A One-Vertex Decomposition Algorithm (1-VDA)}

The algorithm is based on decomposition of the initial graph and all kinds of
subgraphs to new subgraphs.

We conjecture that the representation with the minimum complexity of $Ex(SR)$
derived by 1-VDM is generated when the decomposition vertex is a middle vertex
in the basic group of the $SR$. That is, the number $i$ of the decomposition
vertex for a current $SR$ subgraph which is positioned between vertices $p$
and $q$ is chosen as $\frac{q+p}{2}$ ($\left\lceil \frac{q+p}{2}\right\rceil $
or $\left\lfloor \frac{q+p}{2}\right\rfloor $). In other words, $i$ is equal
to $\frac{q+p}{2}$ for odd $q-p+1$ and $i$ equals $\frac{q+p-1}{2}$ or
$\frac{q+p+1}{2}$ for even $q-p+1$.

\begin{figure}[t]
\setlength{\unitlength}{0.7cm}
\par
\begin{picture}(10,3.3)(-0.5,-1.9)\thicklines
%
\multiput(0,0)(3,0){3}{\circle*{0.1875}}
\multiput(1.5,1.5)(3,0){3}{\circle*{0.1875}}
\multiput(1.5,-1.5)(3,0){2}{%
\circle*{0.1875}}
\put(0,0.35){\makebox(0,0){1}}
\put(3.02,0.35){%
\makebox(0,0){2}}
\put(6,0.35){\makebox(0,0){3}}
\put(1.5,1.8){%
\makebox(0,0){1}}
\put(4.5,1.8){\makebox(0,0){2}}
\put(7.5,1.8){%
\makebox(0,0){3}}
\put(1.5,-1.8){\makebox(0,0){1}}
\put(4.5,-1.8){%
\makebox(0,0){2}}
\multiput(0,0)(3,0){3}{\vector(1,1){1.5}}
\multiput(0,0)(3,0){2}{\vector(1,-1){1.5}}
\multiput(1.5,1.5)(3,0){2}{%
\vector(1,-1){1.5}}
\multiput(1.5,-1.5)(3,0){2}{\vector(1,1){1.5}}
\multiput(0,0)(3,0){2}{\vector(1,0){3}}
\multiput(1.5,1.5)(3,0){2}{%
\vector(1,0){3}}
\multiput(1.5,-1.5)(3,0){1}{\vector(1,0){3}}
\put(1.5,0.3){\makebox(0,0){$b_{1}$}}
\put(4.5,0.3){%
\makebox(0,0){$b_{2}$}}
\put(2.7,1.75){\makebox(0,0){$c_{1}$}}
\put(6,1.75){\makebox(0,0){$c_{2}$}}
\put(2.7,-1.75){%
\makebox(0,0){$a_{1}$}}
\put(0.55,0.95){\makebox(0,0){$e_{1}$}}
\put(3.55,0.95){\makebox(0,0){$e_{3}$}}
\put(6.55,0.95){%
\makebox(0,0){$e_{5}$}}
\put(2.4,0.95){\makebox(0,0){$e_{2}$}}
\put(5.4,0.95){\makebox(0,0){$e_{4}$}}
\put(0.6,-0.95){%
\makebox(0,0){$d_{1}$}}
\put(3.6,-0.95){\makebox(0,0){$d_{3}$}}
\put(2.4,-0.95){\makebox(0,0){$d_{2}$}}
\put(5.4,-0.95){%
\makebox(0,0){$d_{4}$}}
\dashline[75]{0.15}(3,2.5)(3,0.55)
\dashline[75]{0.15}(3,0.1)(3,-2.5)
\end{picture}
\par
\begin{picture}(10,0)(-8.7,-2.5)\thicklines
%
\multiput(0,0)(3,0){4}{\circle*{0.1875}}
\multiput(1.5,1.5)(3,0){4}{\circle*{0.1875}}
\multiput(1.5,-1.5)(3,0){3}{%
\circle*{0.1875}}
\put(0,0.35){\makebox(0,0){1}}
\put(3.02,0.35){%
\makebox(0,0){2}}
\put(6,0.35){\makebox(0,0){3}}
\put(9,0.35){%
\makebox(0,0){4}}
\put(1.5,1.8){\makebox(0,0){1}}
\put(4.5,1.8){%
\makebox(0,0){2}}
\put(7.5,1.8){\makebox(0,0){3}}
\put(10.5,1.8){%
\makebox(0,0){4}}
\put(1.5,-1.8){\makebox(0,0){1}}
\put(4.5,-1.8){%
\makebox(0,0){2}}
\put(7.5,-1.8){\makebox(0,0){3}}
\multiput(0,0)(3,0){4}{\vector(1,1){1.5}}
\multiput(0,0)(3,0){3}{%
\vector(1,-1){1.5}}
\multiput(1.5,1.5)(3,0){3}{\vector(1,-1){1.5}}
\multiput(1.5,-1.5)(3,0){3}{\vector(1,1){1.5}}
\multiput(0,0)(3,0){3}{%
\vector(1,0){3}}
\multiput(1.5,1.5)(3,0){3}{\vector(1,0){3}}
\multiput(1.5,-1.5)(3,0){2}{\vector(1,0){3}}
\put(1.5,0.3){%
\makebox(0,0){$b_{1}$}}
\put(4.5,0.3){\makebox(0,0){$b_{2}$}}
\put(7.5,0.3){\makebox(0,0){$b_{3}$}}
\put(3,1.75){%
\makebox(0,0){$c_{1}$}}
\put(5.7,1.75){\makebox(0,0){$c_{2}$}}
\put(9,1.75){\makebox(0,0){$c_{3}$}}
\put(3,-1.75){%
\makebox(0,0){$a_{1}$}}
\put(5.7,-1.75){\makebox(0,0){$a_{2}$}}
\put(0.55,0.95){\makebox(0,0){$e_{1}$}}
\put(3.55,0.95){%
\makebox(0,0){$e_{3}$}}
\put(6.55,0.95){\makebox(0,0){$e_{5}$}}
\put(9.55,0.95){\makebox(0,0){$e_{7}$}}
\put(2.4,0.95){%
\makebox(0,0){$e_{2}$}}
\put(5.4,0.95){\makebox(0,0){$e_{4}$}}
\put(8.4,0.95){\makebox(0,0){$e_{6}$}}
\put(0.6,-0.95){%
\makebox(0,0){$d_{1}$}}
\put(3.6,-0.95){\makebox(0,0){$d_{3}$}}
\put(6.6,-0.95){\makebox(0,0){$d_{5}$}}
\put(2.4,-0.95){%
\makebox(0,0){$d_{2}$}}
\put(5.4,-0.95){\makebox(0,0){$d_{4}$}}
\put(8.4,-0.95){\makebox(0,0){$d_{6}$}}
\dashline[75]{0.15}(6,2.5)(6,0.55)
\dashline[75]{0.15}(6,0.1)(6,-2.5)
\end{picture}\caption{Decomposition of single-leaf square rhomboids by 1-VDA.}%
\label{rhom_fig13}%
\end{figure}
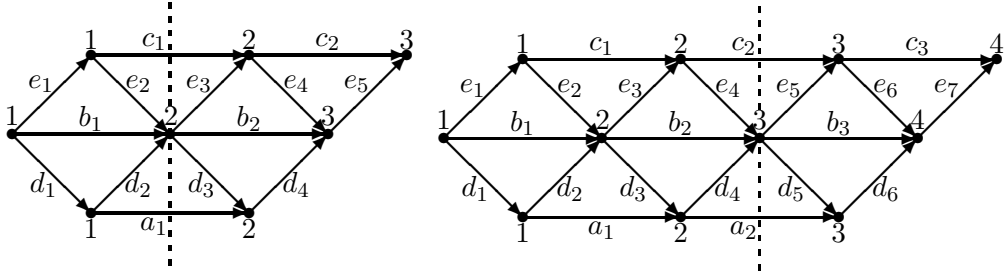

An $\widehat{SR}$ subgraph is decomposed through a decomposition vertex
selected in its basic group into six new subgraphs in the same way as an $SR$
(see the examples in Figure \ref{rhom_fig13}). The decomposition vertex is
chosen so that the location of the split is in the middle of the subgraph.

\begin{figure}[ptbh]
\setlength{\unitlength}{0.7cm}
\par
\begin{picture}(10,2)(5,0.4)\thicklines
\put(5.3,0){\makebox(0,0){(a)}}
\multiput(6,0)(3,0){2}{\circle*{0.1875}}
\multiput(4.5,1.5)(3,0){3}{\circle*{0.1875}}
\multiput(7.5,-1.5)(3,0){1}{\circle*{0.1875}}
\put(6,0.35){\makebox(0,0){6}} \put(9,0.35){\makebox(0,0){7}}
\put(4.5,1.8){\makebox(0,0){5}} \put(7.5,1.8){\makebox(0,0){6}}
\put(10.5,1.8){\makebox(0,0){7}}
\put(7.5,-1.8){\makebox(0,0){6}}
\multiput(6,0)(3,0){2}{\vector(1,1){1.5}}
\multiput(6,0)(3,0){1}{\vector(1,-1){1.5}}
\multiput(4.5,1.5)(3,0){2}{\vector(1,-1){1.5}}
\multiput(7.5,-1.5)(3,0){1}{\vector(1,1){1.5}}
\multiput(6,0)(3,0){1}{\vector(1,0){3}}
\multiput(4.5,1.5)(3,0){2}{\vector(1,0){3}}
\put(7.5,0.3){\makebox(0,0){$b_{6}$}}
\put(6,1.75){\makebox(0,0){$c_{5}$}}
\put(9,1.75){\makebox(0,0){$c_{6}$}}
\put(6.55,0.95){\makebox(0,0){$e_{11}$}}
\put(9.55,0.95){\makebox(0,0){$e_{13}$}}
\put(5.4,0.95){\makebox(0,0){$e_{10}$}}
\put(8.4,0.95){\makebox(0,0){$e_{12}$}}
\put(6.6,-0.95){\makebox(0,0){$d_{11}$}}
\put(8.4,-0.95){\makebox(0,0){$d_{12}$}}
\end{picture}
\par
\begin{picture}(10,1)(0,-0.6)\thicklines
\put(5.3,0){\makebox(0,0){(b)}}
\multiput(6,0)(3,0){2}{\circle*{0.1875}}
\multiput(7.5,1.5)(3,0){2}{\circle*{0.1875}}
\multiput(4.5,-1.5)(3,0){2}{\circle*{0.1875}}
\put(6,0.35){\makebox(0,0){6}} \put(9,0.35){\makebox(0,0){7}}
\put(7.5,1.8){\makebox(0,0){6}} \put(10.5,1.8){\makebox(0,0){7}}
\put(4.5,-1.8){\makebox(0,0){5}} \put(7.5,-1.8){\makebox(0,0){6}}
\multiput(6,0)(3,0){2}{\vector(1,1){1.5}}
\multiput(6,0)(3,0){1}{\vector(1,-1){1.5}}
\multiput(7.5,1.5)(3,0){1}{\vector(1,-1){1.5}}
\multiput(4.5,-1.5)(3,0){2}{\vector(1,1){1.5}}
\multiput(6,0)(3,0){1}{\vector(1,0){3}}
\multiput(7.5,1.5)(3,0){1}{\vector(1,0){3}}
\multiput(4.5,-1.5)(3,0){1}{\vector(1,0){3}}
\put(7.5,0.3){\makebox(0,0){$b_{6}$}}
\put(9,1.75){\makebox(0,0){$c_{6}$}}
\put(6,-1.75){\makebox(0,0){$a_{5}$}}
\put(6.55,0.95){\makebox(0,0){$e_{11}$}}
\put(9.55,0.95){\makebox(0,0){$e_{13}$}}
\put(8.4,0.95){\makebox(0,0){$e_{12}$}}
\put(6.6,-0.95){\makebox(0,0){$d_{11}$}}
\put(5.4,-0.95){\makebox(0,0){$d_{10}$}}
\put(8.4,-0.95){\makebox(0,0){$d_{12}$}}
\end{picture}
\par
\begin{picture}(10,0)(-10,-1.2)\thicklines
\put(5.3,0){\makebox(0,0){(d)}}
\multiput(6,0)(3,0){2}{\circle*{0.1875}}
\multiput(4.5,1.5)(3,0){2}{\circle*{0.1875}}
\multiput(7.5,-1.5)(3,0){2}{\circle*{0.1875}}
\put(6,0.35){\makebox(0,0){6}} \put(9,0.35){\makebox(0,0){7}}
\put(4.5,1.8){\makebox(0,0){5}} \put(7.5,1.8){\makebox(0,0){6}}
\put(7.5,-1.8){\makebox(0,0){6}} \put(10.5,-1.8){\makebox(0,0){7}}
\multiput(6,0)(3,0){1}{\vector(1,1){1.5}}
\multiput(6,0)(3,0){2}{\vector(1,-1){1.5}}
\multiput(4.5,1.5)(3,0){2}{\vector(1,-1){1.5}}
\multiput(7.5,-1.5)(3,0){1}{\vector(1,1){1.5}}
\multiput(6,0)(3,0){1}{\vector(1,0){3}}
\multiput(4.5,1.5)(3,0){1}{\vector(1,0){3}}
\multiput(7.5,-1.5)(3,0){1}{\vector(1,0){3}}
\put(7.5,0.3){\makebox(0,0){$b_{6}$}}
\put(6,1.75){\makebox(0,0){$c_{5}$}}
\put(9,-1.75){\makebox(0,0){$a_{6}$}}
\put(6.55,0.95){\makebox(0,0){$e_{11}$}}
\put(5.4,0.95){\makebox(0,0){$e_{10}$}}
\put(8.4,0.95){\makebox(0,0){$e_{12}$}}
\put(6.6,-0.95){\makebox(0,0){$d_{11}$}}
\put(9.6,-0.95){\makebox(0,0){$d_{13}$}}
\put(8.4,-0.95){\makebox(0,0){$d_{12}$}}
\end{picture}
\par
\begin{picture}(10,0)(-5,-1.9)\thicklines
\put(5.3,0){\makebox(0,0){(c)}}
\multiput(6,0)(3,0){2}{\circle*{0.1875}}
\multiput(7.5,1.5)(3,0){1}{\circle*{0.1875}}
\multiput(4.5,-1.5)(3,0){3}{\circle*{0.1875}}
\put(6,0.35){\makebox(0,0){6}} \put(9,0.35){\makebox(0,0){7}}
\put(7.5,1.8){\makebox(0,0){6}}
\put(4.5,-1.8){\makebox(0,0){5}} \put(7.5,-1.8){\makebox(0,0){6}}
\put(10.5,-1.8){\makebox(0,0){7}}
\multiput(6,0)(3,0){1}{\vector(1,1){1.5}}
\multiput(6,0)(3,0){2}{\vector(1,-1){1.5}}
\multiput(7.5,1.5)(3,0){1}{\vector(1,-1){1.5}}
\multiput(4.5,-1.5)(3,0){2}{\vector(1,1){1.5}}
\multiput(6,0)(3,0){1}{\vector(1,0){3}}
\multiput(4.5,-1.5)(3,0){2}{\vector(1,0){3}}
\put(7.5,0.3){\makebox(0,0){$b_{6}$}}
\put(6,-1.75){\makebox(0,0){$a_{5}$}}
\put(9,-1.75){\makebox(0,0){$a_{6}$}}
\put(6.55,0.95){\makebox(0,0){$e_{11}$}}
\put(8.4,0.95){\makebox(0,0){$e_{12}$}}
\put(6.6,-0.95){\makebox(0,0){$d_{11}$}}
\put(9.6,-0.95){\makebox(0,0){$d_{13}$}}
\put(5.4,-0.95){\makebox(0,0){$d_{10}$}}
\put(8.4,-0.95){\makebox(0,0){$d_{12}$}}
\end{picture}\caption{Varieties of a dipterous square rhomboid of size $2$. }%
\label{rhom_fig6}%
\end{figure}
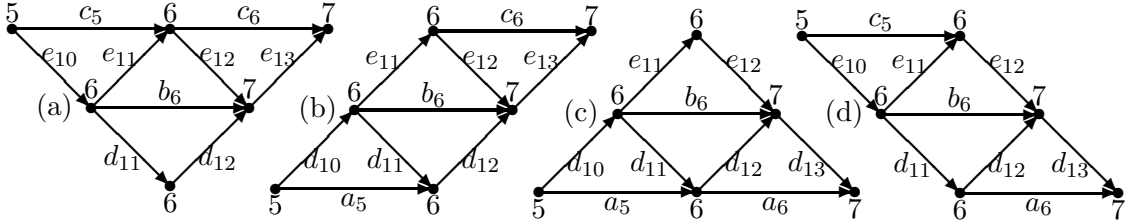

Three kinds of subgraphs are revealed in an $\widehat{SR}$ in the course of
decomposition. The first and the second of them are an $SR$ and an
$\widehat{SR}$, respectively. The third one is an $SR$ supplemented by two
additional pairs of edges (one pair is on the left and another one is on the
right). Possible varieties of this st-dag (we call it a\textit{\ dipterous
square rhomboid} and denote it by $\widehat{\widehat{SR}}$) are illustrated in
Figure \ref{rhom_fig6}(a, b, c, d). We define subgraphs illustrated in Figure
\ref{rhom_fig6}(a, c) as \textit{trapezoidal} $\widehat{\widehat{SR}}$
\textit{graphs} and subgraphs illustrated in Figure \ref{rhom_fig6}(b, d) as
\textit{parallelogram} $\widehat{\widehat{SR}}$ \textit{graphs}. Let
$\widehat{\widehat{SR}}(n)$ (an $\widehat{\widehat{SR}}$ of size $n$) denote
an $\widehat{\widehat{SR}}$ including $n$ basic vertices.

We denote by $E(\overline{p},\overline{q})$, $E(\underline{p},\overline{q})$,
$E(\underline{p},\underline{q})$, $E(\overline{p},\underline{q})$
subexpressions related to subgraphs of Figure \ref{rhom_fig6}(a, b, c, d,
respectively) with a source $\overline{p}$ and a sink $\overline{q}$, a source
$\underline{p}$ and a sink $\overline{q}$, a source $\underline{p}$ and a sink
$\underline{q}$, and a source $\overline{p}$ and a sink $\underline{q}%
$.

\begin{figure}[ptbh]
\setlength{\unitlength}{0.7cm}
\par
\begin{picture}(10,3)(-4,-1.5)\thicklines
%
\multiput(9,0)(3,0){3}{%
\circle*{0.1875}}
\multiput(10.5,1.5)(3,0){3}{\circle*{0.1875}}
\multiput(7.5,-1.5)(3,0){3}{\circle*{0.1875}}
\put(9,0.35){\makebox(0,0){5}}
\put(12,0.35){\makebox(0,0){6}}
\put(15,0.35){\makebox(0,0){7}}
\put(10.5,1.8){\makebox(0,0){5}}
\put(13.5,1.8){\makebox(0,0){6}}
\put(16.5,1.8){\makebox(0,0){7}}
\put(7.5,-1.8){\makebox(0,0){4}}
\put(10.5,-1.8){\makebox(0,0){5}}
\put(13.5,-1.8){\makebox(0,0){6}}
\multiput(9,0)(3,0){3}{\vector(1,1){1.5}}
\multiput(9,0)(3,0){2}{\vector(1,-1){1.5}}
\multiput(10.5,1.5)(3,0){2}{%
\vector(1,-1){1.5}}
\multiput(7.5,-1.5)(3,0){3}{\vector(1,1){1.5}}
\multiput(9,0)(3,0){2}{\vector(1,0){3}}
\multiput(10.5,1.5)(3,0){2}{%
\vector(1,0){3}}
\multiput(7.5,-1.5)(3,0){2}{\vector(1,0){3}}
\put(10.5,0.3){\makebox(0,0){$b_{5}$}}
\put(13.5,0.3){\makebox(0,0){$b_{6}$}}
\put(11.7,1.75){\makebox(0,0){$c_{5}$}}
\put(15,1.75){\makebox(0,0){$c_{6}$}}
\put(9,-1.75){\makebox(0,0){$a_{4}$}}
\put(11.7,-1.75){\makebox(0,0){$a_{5}$}}
\put(9.55,0.95){\makebox(0,0){$e_{9}$}}
\put(12.55,0.95){\makebox(0,0){$e_{11}$}}
\put(15.5,0.95){\makebox(0,0){$e_{13}$}}
\put(11.4,0.95){\makebox(0,0){$e_{10}$}}
\put(14.45,0.95){\makebox(0,0){$e_{12}$}}
\put(9.6,-0.95){\makebox(0,0){$d_{9}$}}
\put(12.6,-0.95){\makebox(0,0){$d_{11}$}}
\put(8.4,-0.95){\makebox(0,0){$d_{8}$}}
\put(11.4,-0.95){\makebox(0,0){$d_{10}$}}
\put(14.45,-0.95){\makebox(0,0){$d_{12}$}}
\dashline[75]{0.15}(12,2.5)(12,0.55)
\dashline[75]{0.15}(12,0.1)(12,-2.5)
\end{picture}
\par
\begin{picture}(10,0)(10.7,-2.2)\thicklines
%
\multiput(12,0)(3,0){4}{\circle*{0.1875}}
\multiput(10.5,1.5)(3,0){5}{\circle*{0.1875}}
\multiput(13.5,-1.5)(3,0){3}{%
\circle*{0.1875}}
\put(12,0.35){\makebox(0,0){6}}
\put(15,0.35){%
\makebox(0,0){7}}
\put(18,0.35){\makebox(0,0){8}}
\put(21.02,0.35){%
\makebox(0,0){9}}
\put(10.5,1.8){\makebox(0,0){5}}
\put(13.5,1.8){%
\makebox(0,0){6}}
\put(16.5,1.8){\makebox(0,0){7}}
\put(19.5,1.8){%
\makebox(0,0){8}}
\put(22.5,1.8){\makebox(0,0){9}}
\put(13.5,-1.8){%
\makebox(0,0){6}}
\put(16.5,-1.8){\makebox(0,0){7}}
\put(19.5,-1.8){%
\makebox(0,0){8}}
\multiput(12,0)(3,0){4}{\vector(1,1){1.5}}
\multiput(12,0)(3,0){3}{\vector(1,-1){1.5}}
\multiput(10.5,1.5)(3,0){4}{%
\vector(1,-1){1.5}}
\multiput(13.5,-1.5)(3,0){3}{\vector(1,1){1.5}}
\multiput(12,0)(3,0){3}{\vector(1,0){3}}
\multiput(10.5,1.5)(3,0){4}{%
\vector(1,0){3}}
\multiput(13.5,-1.5)(3,0){2}{\vector(1,0){3}}
\put(13.5,0.3){\makebox(0,0){$b_{6}$}}
\put(16.5,0.3){%
\makebox(0,0){$b_{7}$}}
\put(19.5,0.3){\makebox(0,0){$b_{8}$}}
\put(12,1.75){\makebox(0,0){$c_{5}$}}
\put(15,1.75){%
\makebox(0,0){$c_{6}$}}
\put(17.7,1.75){\makebox(0,0){$c_{7}$}}
\put(21,1.75){\makebox(0,0){$c_{8}$}}
\put(15,-1.75){%
\makebox(0,0){$a_{6}$}}
\put(17.7,-1.75){\makebox(0,0){$a_{7}$}}
\put(12.55,0.95){\makebox(0,0){$e_{11}$}}
\put(15.5,0.95){%
\makebox(0,0){$e_{13}$}}
\put(18.5,0.95){\makebox(0,0){$e_{15}$}}
\put(21.5,0.95){\makebox(0,0){$e_{17}$}}
\put(11.45,0.95){%
\makebox(0,0){$e_{10}$}}
\put(14.45,0.95){%
\makebox(0,0){$e_{12}$}}
\put(17.45,0.95){\makebox(0,0){$e_{14}$}}
\put(20.45,0.95){\makebox(0,0){$e_{16}$}}
\put(12.6,-0.95){%
\makebox(0,0){$d_{11}$}}
\put(15.6,-0.95){\makebox(0,0){$d_{13}$}}
\put(18.6,-0.95){\makebox(0,0){$d_{15}$}}
\put(14.45,-0.95){%
\makebox(0,0){$d_{12}$}}
\put(17.45,-0.95){\makebox(0,0){$d_{14}$}}
\put(20.45,-0.95){\makebox(0,0){$d_{16}$}}
\dashline[75]{0.15}(18,2.5)(18,0.55)
\dashline[75]{0.15}(18,0.1)(18,-2.5)
\end{picture}\caption{Decomposition of dipterous square rhomboids by 1-VDA.}%
\label{rhom_fig14}%
\end{figure}
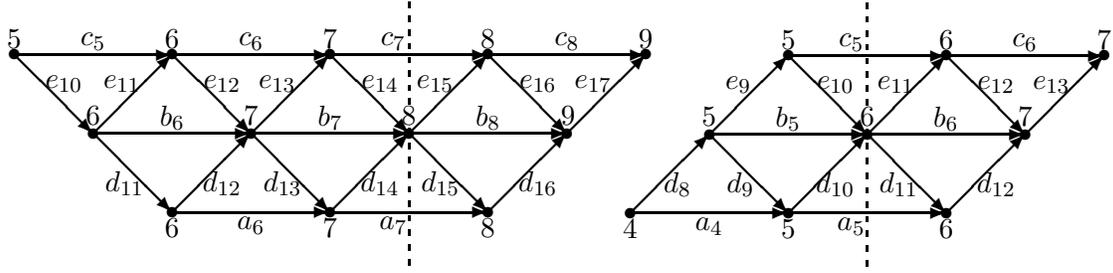

An $\widehat{\widehat{SR}}$ subgraph is decomposed into six new subgraphs in
the same way as an $SR$ and an $\widehat{SR}$ (see the examples in Figure
\ref{rhom_fig14}). The number $i$ of the decomposition vertex in the basic
group for a current $\widehat{\widehat{SR}}$ subgraph which is positioned
between vertices $p$ and $q$, is chosen as $\frac{q+p+1}{2}$ ($\left\lceil
\frac{q+p+1}{2}\right\rceil $ or $\left\lfloor \frac{q+p+1}{2}\right\rfloor
$).

In the course of decomposition, two kinds of subgraphs are revealed in an
$\widehat{\widehat{SR}}$. They are an $\widehat{SR}$ and an $\widehat
{\widehat{SR}}$. Therefore, no new kinds of subgraphs are formed and
subexpressions related to all subgraphs can be generated.

Hence, $Ex(SR)$ is computed by the following recursive relations:\bigskip

\label{mfda_1} $\ 1.$ $E(p,p)=1$

$\ 2.$ $E(p,\overline{p})=e_{2p-1}$ \ \ \ $3.$ $E(p,\underline{p})=d_{2p-1}$
\ \ \ $4.$ $E(\overline{p},p+1)=e_{2p}$ \ \ \ $5.$ $E(\underline
{p},p+1)=d_{2p}$

$\ 6.$ $E(\overline{p},\overline{p+1})=c_{p}+e_{2p}e_{2p+1}$ \ \ \ $7.$
$E(\overline{p},\underline{p+1})=e_{2p}d_{2p+1}$

$\ 8.$ $E(\underline{p},\overline{p+1})=d_{2p}e_{2p+1}$ \ \ \ $9.$
$E(\underline{p},\underline{p+1})=a_{p}+d_{2p}d_{2p+1}$

$10.$ $E(p,p+1)=b_{p}+e_{2p-1}e_{2p}+d_{2p-1}d_{2p}$

$11.$ \label{mfda_11}$E(p,\overline{p+1})=(b_{p}+d_{2p-1}d_{2p})e_{2p+1}%
+e_{2p-1}(c_{p}+e_{2p}e_{2p+1})$

$12.$ $E(p,\underline{p+1})=(b_{p}+e_{2p-1}e_{2p})d_{2p+1}+d_{2p-1}%
(a_{p}+d_{2p}d_{2p+1})$

$13.$ $E(\overline{p},p+2)=(c_{p}+e_{2p}e_{2p+1})e_{2p+2}+e_{2p}%
(b_{p+1}+d_{2p+1}d_{2p+2})$

$14.$ $E(\underline{p},p+2)=(a_{p}+d_{2p}d_{2p+1})d_{2p+2}+d_{2p}%
(b_{p+1}+e_{2p+1}e_{2p+2})$

$15.$ $E(\overline{p},\overline{p+2})=e_{2p}(b_{p+1}+d_{2p+1}d_{2p+2}%
)e_{2p+3}+(a_{p}+d_{2p}d_{2p+1})(a_{p+1}+d_{2p+2}d_{2p+3})$

$16.$ $E(\overline{p},\underline{p+2})=e_{2p}(b_{p+1}d_{2p+3}+d_{2p+1}%
(a_{p+1}+d_{2p+2}d_{2p+3}))+(c_{p}+e_{2p}e_{2p+1})e_{2p+2}d_{2p+3}$

$17.$ $E(\underline{p},\overline{p+2})=d_{2p}(b_{p+1}e_{2p+3}+e_{2p+1}%
(c_{p+1}+e_{2p+2}e_{2p+3}))+(a_{p}+d_{2p}d_{2p+1})d_{2p+2}e_{2p+3}$

$18.$ \label{mfda_18}$E(\underline{p},\underline{p+2})=d_{2p}(b_{p+1}%
+e_{2p+1}e_{2p+2})d_{2p+3}+(c_{p}+e_{2p}e_{2p+1})(c_{p+1}+e_{2p+2}e_{2p+3})$

$19.$ \label{mfda_19}$E(p,\overline{q})=E(p,i)E(i,\overline{q})+E(p,\overline
{i-1})c_{i-1}E(\overline{i},\overline{q})+E(p,\underline{i-1})a_{i-1}%
E(\underline{i},\overline{q})$, $i=\left\lceil \frac{q+p}{2}\right\rceil $

\quad$\ (q>p+1)$

$20.$ $E(p,\underline{q})=E(p,i)E(i,\underline{q})+E(p,\overline{i-1}%
)c_{i-1}E(\overline{i},\underline{q})+E(p,\underline{i-1})a_{i-1}%
E(\underline{i},\underline{q})$, $i=\left\lceil \frac{q+p}{2}\right\rceil $

\quad$\ (q>p+1)$

$21.$ $E(\overline{p},q)=E(\overline{p},i)E(i,q)+E(\overline{p},\overline
{i-1})c_{i-1}E(\overline{i},q)+E(\overline{p},\underline{i-1})a_{i-1}%
E(\underline{i},q)$, $i=\left\lceil \frac{q+p}{2}\right\rceil $

\quad$\ (q>p+2)$

$22.$ $E(\underline{p},q)=E(\underline{p},i)E(i,q)+E(\underline{p}%
,\overline{i-1})c_{i-1}E(\overline{i},q)+E(\underline{p},\underline
{i-1})a_{i-1}E(\underline{i},q)$, $i=\left\lceil \frac{q+p}{2}\right\rceil $

\quad$\ (q>p+2)$

$23.$ \label{mfda_23}$E(\overline{p},\overline{q})=E(\overline{p}%
,i)E(i,\overline{q})+E(\overline{p},\overline{i-1})c_{i-1}E(\overline
{i},\overline{q})+E(\overline{p},\underline{i-1})a_{i-1}E(\underline
{i},\overline{q})$, $i=\frac{q+p+1}{2}$

\quad$\ (q>p+2)$

$24.$ $E(\overline{p},\underline{q})=E(\overline{p},i)E(i,\underline
{q})+E(\overline{p},\overline{i-1})c_{i-1}E(\overline{i},\underline
{q})+E(\overline{p},\underline{i-1})a_{i-1}E(\underline{i},\underline{q})$,
$i=\frac{q+p+1}{2}$

\quad$\ (q>p+2)$

$25.$ $E(\underline{p},\overline{q})=E(\underline{p},i)E(i,\overline
{q})+E(\underline{p},\overline{i-1})c_{i-1}E(\overline{i},\overline
{q})+E(\underline{p},\underline{i-1})a_{i-1}E(\underline{i},\overline{q})$,
$i=\frac{q+p+1}{2}$

\quad$\ (q>p+2)$

$26.$ \label{mfda_26}$E(\underline{p},\underline{q})=E(\underline
{p},i)E(i,\underline{q})+E(\underline{p},\overline{i-1})c_{i-1}E(\overline
{i},\underline{q})+E(\underline{p},\underline{i-1})a_{i-1}E(\underline
{i},\underline{q})$, $i=\frac{q+p+1}{2}$

\quad$\ (q>p+2)$

$27.$ \label{mfda_27}$E(p,q)=E(p,i)E(i,q)+E(p,\overline{i-1})c_{i-1}%
E(\overline{i},q)+E(p,\underline{i-1})a_{i-1}E(\underline{i},q)$,
$i=\frac{q+p}{2}$

\quad$\ (q>p+1)$.

Subgraphs of sizes $1$ and $2$ are trivial and their expressions are in lines
(1 -- 18). Specifically, expressions of subgraphs $\widehat{SR}(2)$ and
$\widehat{\widehat{SR}}(2)$ are presented in the minimum factored form (lines
11 -- 18).

For example, the expression of the square rhomboid of size $3$ derived by
1-VDA is%

\[
(b_{1}+e_{1}e_{2}+d_{1}d_{2})(b_{2}+e_{3}e_{4}+d_{3}d_{4})+e_{1}c_{1}%
e_{4}+d_{1}a_{1}d_{4}\text{.}%
\]
It contains $16$ literals.

The following lemma results from lines (23 -- 26) of 1-VDA.

\begin{lemma}
\label{lem_mfda}Complexities of expressions $Ex\left(  trapezoidal\text{
}\widehat{\widehat{SR}}(n)\right)  $ and\linebreak\ $Ex\left(
parallelogram\text{ }\widehat{\widehat{SR}}(n)\right)  $ derived by 1-VDA are
equal for $n>2$.
\end{lemma}

The following proposition results from Lemma \ref{lem_mfda} and from lines (1
-- 27) of 1-VDA.

\begin{proposition}
\label{th_mfda}The total number of literals $T(n)$ in the expression
$Ex(SR(n))$ derived by 1-VDA is defined recursively as follows:

1) $T(1)=0;\ $2) $\widehat{T}(1)=1;\ $3) $\widehat{\widehat{T}}_{pr}%
(1)=2;\ $4) $\widehat{\widehat{T}}_{tr}(1)=3$

5) $T(2)=5;$ 6)$\widehat{T}(2)=8;\ $7) $\widehat{\widehat{T}}_{pr}(2)=12;\ $8)
$\widehat{\widehat{T}}_{tr}(2)=11$

9) $\widehat{T}(3)=22;\ $10) $\widehat{\widehat{T}}(3)=28;$ 11) $\widehat
{T}(4)=47;\ $12) $\widehat{\widehat{T}}(4)=60$

13) $\widehat{T}(5)=79;\ $14) $\widehat{\widehat{T}}(5)=92;$ 15) $\widehat
{T}(6)=132;\ $16) $\widehat{\widehat{T}}(6)=50$

17) $T(n)=T\left(  \left\lceil \frac{n}{2}\right\rceil \right)  +T\left(
\left\lfloor \frac{n}{2}\right\rfloor +1\right)  +2\widehat{T}\left(
\left\lceil \frac{n}{2}\right\rceil -1\right)  +2\widehat{T}\left(
\left\lfloor \frac{n}{2}\right\rfloor \right)  +2\quad(n>\nolinebreak2)$

18) $\widehat{T}(n)=T\left(  \left\lfloor \frac{n}{2}\right\rfloor +1\right)
+\widehat{T}\left(  \left\lceil \frac{n}{2}\right\rceil \right)  +2\widehat
{T}\left(  \left\lfloor \frac{n}{2}\right\rfloor \right)  +2\widehat
{\widehat{T}}\left(  \left\lceil \frac{n}{2}\right\rceil -1\right)
+2\quad(n>\nolinebreak6)$

19) $\widehat{\widehat{T}}(n)=\widehat{T}\left(  \left\lceil \frac{n}%
{2}\right\rceil \right)  +\widehat{T}\left(  \left\lfloor \frac{n}%
{2}\right\rfloor +1\right)  +2\widehat{\widehat{T}}\left(  \left\lceil
\frac{n}{2}\right\rceil -1\right)  +2\widehat{\widehat{T}}\left(  \left\lfloor
\frac{n}{2}\right\rfloor \right)  +2\quad(n>\nolinebreak6)$,\smallskip

where $\widehat{T}(n)$, $\widehat{\widehat{T}}_{pr}(1)$, $\widehat{\widehat
{T}}_{tr}(1)$, and $\widehat{\widehat{T}}(n)$ are the total numbers of
literals in $Ex(\widehat{SR}(n))$, $Ex\left(  parallelogram\text{ }%
\widehat{\widehat{SR}}\left(  1\right)  \right)  $, $Ex\left(
trapezoidal\text{ }\widehat{\widehat{SR}}\left(  1\right)  \right)  $, and
$Ex\left(  \widehat{\widehat{SR}}\left(  n\right)  \right)  $, respectively.
\end{proposition}

Hence, the expression $Ex(SR(n))$ derived by 1-VDA consists of six
subexpressions related to six revealed subgraphs of size $n^{\prime}\approx$
$\frac{n}{2}$ and two additional literals. Each of these subexpressions is
constructed in its turn from six subexpressions related to six subgraphs of
size $n^{\prime\prime}\approx$ $\frac{n}{4}$ and two additional literals, etc.
Thus, by the master theorem, $T(n)=O\left(  n^{\log_{2}6}\right)  $, i.e.,
1-VDA provides the representation of $Ex(SR(n))$ with a polynomial complexity.

It is of interest to obtain exact formulae describing complexity of the
expression $Ex(SR(n))$ derived by 1-VDA. We attempt to do it for $n$ that is a
power of two, i.e., $n=2^{k}$ for some positive integer $k\geq2$. Statements
(17 -- 19) of Proposition \ref{th_mfda} are presented for $n=2^{k} $ $\left(
k\geq3\right)  $ as
\begin{equation}
\left\{
\begin{array}
[c]{l}%
T\left(  n\right)  =T\left(  \frac{n}{2}+1\right)  +T\left(  \frac{n}%
{2}\right)  +2\widehat{T}\left(  \frac{n}{2}\right)  +2\widehat{T}\left(
\frac{n}{2}-1\right)  +2\\
\widehat{T}\left(  n\right)  =T\left(  \frac{n}{2}+1\right)  +3\widehat
{T}\left(  \frac{n}{2}\right)  +2\widehat{\widehat{T}}\left(  \frac{n}%
{2}-1\right)  +2\\
\widehat{\widehat{T}}\left(  n\right)  =\widehat{T}\left(  \frac{n}%
{2}+1\right)  +\widehat{T}\left(  \frac{n}{2}\right)  +2\widehat{\widehat{T}%
}\left(  \frac{n}{2}\right)  +2\widehat{\widehat{T}}\left(  \frac{n}%
{2}-1\right)  +2\text{ ,}%
\end{array}
\right.  \label{rhomf3}%
\end{equation}
respectively.

After a number of transformations, the following explicit formulae for
simultaneous recurrences (\ref{rhomf3})\textbf{\ }are obtained by the method
for linear recurrence relations solving \cite{Ros}:
\begin{align}
T(n)  &  =\frac{154}{135}n^{\log_{2}6}+\frac{1}{27}n^{\log_{2}3}-\frac{2}%
{5}\nonumber\\
\widehat{T}(n)  &  =\frac{154}{135}n^{\log_{2}6}+\frac{19}{27}n^{\log_{2}%
3}-\frac{2}{5}\label{rhomf28}\\
\widehat{\widehat{T}}(n)  &  =\frac{154}{135}n^{\log_{2}6}+\frac{58}%
{27}n^{\log_{2}3}-\frac{2}{5}\text{.}\nonumber
\end{align}

\section{Comparison of 1-VDA with Other Algorithms}

Another decomposition method (called 2-VDM) for generating algebraic
expressions of square rhomboids is presented in \cite{KoL2} and \cite{KoL3}.
The method consists in splitting a square rhomboid through two decomposition
vertices one of which belongs to the upper group and another one to the lower
group. The following algorithms based on this method are considered in
\cite{KoL2} and \cite{KoL3}: full decomposition algorithm (FDA), combined
decomposition algorithm (CDA) and improved FDA (IFDA). Complexities of
generated by these algorithms expressions are also $O\left(  n^{\log_{2}%
6}\right)  $. The values of $T(n)$ for these algorithms and for 1-VDA are
presented in Table \ref{tab1} ($4\leq n\leq10$).%

\begin{table}[tbph] \centering
\begin{tabular}
[c]{||c||c|c|c|c||}\hline\hline
$n$ & $T(n)$, FDA & $T\left(  n\right)  $, CDA & $T(n)$, IFDA & $T\left(
n\right)  $, 1-VDA\\\hline
$4$ & $47$ & $43$ & $43$ & $41$\\\hline
$5$ & $110$ & $102$ & $100$ & $66$\\\hline
$6$ & $173$ & $161$ & $157$ & $119$\\\hline
$7$ & $252$ & $236$ & $228$ & $172$\\\hline
$8$ & $331$ & $311$ & $299$ & $247$\\\hline
$9$ & $520$ & $488$ & $470$ & $322$\\\hline
$10$ & $709$ & $665$ & $641$ & $439$\\\hline\hline
\end{tabular}
\caption{Complexities for FDA, CDA, IFDA, and
1-VDA.\label{tab1}}%
\end{table}%

One can see that 1-VDA is more efficient than FDA, CDA, and IFDA.

The numerical advantage of 1-VDA in contrast to other algorithms follows also
from presented in \cite{KoL2} and \cite{KoL3} explicit formulae analogous to
(\ref{rhomf28}).The coefficient of the leading term of the formulae -
$n^{\log_{2}6}$ is equal to $\frac{79}{45}\approx1.76$ for FDA, $\frac
{227}{135}\approx1.68$ for CDA, $\frac{212}{135}\approx1.57$ for IFDA, and
$\frac{154}{135}\approx1.14$ for 1-VDA.

\section{Conclusions}

The decomposition method divides all subgraphs, which are revealed in the
current recursive step into a constant number of subgraphs of proportionally
decreasing sizes. The existence of a decomposition method for an st-dag $G$ is
a sufficient condition for the existence of a representation of $Ex\left(
G\right)  $ with a polynomial complexity. The complexity depends on the number
of revealed subgraphs in each recursive step.

\end{document}